\documentclass[prl,twocolumn]{revtex4}

\usepackage{graphicx}

\begin{document}

\title{ Quantum Fluctuation-Induced Uniaxial and Biaxial Spin Nematics}
\author{Jun Liang Song, Gordon W. Semenoff, and Fei Zhou}
\affiliation{
Department of Physics and Astronomy,
The University of British Columbia, Vancouver, B. C., Canada V6T1Z1}
\date{{\small \today}}

\begin{abstract}
It is shown that zero point quantum fluctuations (ZPQFs) completely
lift the accidental continuous degeneracy that is found in mean
field analysis of quantum spin nematic phases of hyperfine spin 2
cold atoms.  The result is two distinct ground states which have
higher symmetries: a uniaxial spin nematic and a biaxial spin
nematic with dihedral symmetry ${Dih}_4$. There is a novel first
order quantum phase transition  between the two phases as atomic
scattering lengths are varied.  We find that the ground state of
$^{87}Rb$ atoms should be a uniaxial spin nematic. We note that
the energy barrier between the phases could be observable in dynamical experiments.
\end{abstract}
\maketitle

Since the successful observation of magnetic ordering in
condensates of sodium atoms in optical traps\cite{Stenger98}, the subject of spinor
gases has attracted much
attention\cite{Ho98,Ohmi98,Ciobanu00,Koashi00,Santos06,Diener06}.
The impressive efforts to measure spin dependent interaction
energies or two-body scattering lengths of cold atoms
\cite{Roberts98,vanKempen02,Schmaljohann04,Widera06,Chang05} and recent progress in cooling
hyperfine spin $F=3$ chromium atoms\cite{Griesmaier05} have
further stimulated extensive interest in spin correlated many-body
states of cold atoms. 

In addition, in recent work, a much better understanding of
spin collective phenomena of cold atoms in optical lattices has been
achieved. Particularly among many other states in optical
lattices\cite{Greiner02}, various quantum spin nematics have received much
attention.
Spin nematics have been suggested as
ground states of $F=1$ sodium atoms ($^{23}Na$), $F=2$ rubidium
atoms ($^{87}Rb$), and $F=3$ chromium atoms ($^{52}Cr$) in optical
lattices \cite{Demler02,Zhou06,Barnett06,Kim06,Eckert06}; they are
also suggested to be relevant to $F=3/2$ cold
atoms\cite{Wu03,Tu06}. Quantum spin nematics are fascinating
because of the rich topology of their ground state manifolds which
gives rise to exotic spin defects and unconventional vortices with
non-integer circulation integrals.
Spin nematics haven't been successfully realized in
conventional solid state systems but it is widely believed that at
least some of them are likely to be created and observed in cold
atomic vapors. The spin nematics of bosonic cold atoms that have
been discussed so far do not break the (lattice) translational symmetry, but
do break rotational symmetry. The absence of lattice translational
symmetry breaking can be attributed to
the relatively weak super exchange interaction between two bosons
if they are spatially antisymmetric.

In fact, our mean field calculations show that for cold atoms such as rubidium 
($^{87}Rb$) in the $F=2$ manifold, a one-parameter family of spin
nematic ground states, including ones which are characterized by
completely different symmetries, are degenerate in energy (see discussions after Eq.(\ref{energy})). In
particular, uniaxial nematics that are invariant under any
rotation around an easy axis are degenerate with biaxial nematics
which are invariant only under a dihedral group ( either $Dih_2$
or $Dih_4$ depending on eigenvalues of the nematic matrix, see
below). The origin of this degeneracy is accidental 
due to the special form of energy in the dilute limit, 
very analogous to the highly degenerate family
of solutions for d-wave pairing obtained by Mermin in the
1970's\cite{Mermin74}.
In this Letter we shall study the effect of zero point quantum
fluctuations (ZPQFs) on nematic order in cold atoms. ZPQFs are
known to lead to a wide range of physical phenomena, including the
Lamb shift in atomic physics\cite{Lamb47}, the Coleman-Weinberg
mechanism of spontaneous symmetry breaking\cite{Coleman74},
fluctuation-induced first order transitions in liquid crystals and
superconductors\cite{Halperin74}, and {\em order from disorder} in
magnetic systems\cite{Henley89,Shender82}. For $F=2$ cold atoms,
we show that ZPQFs completely lift the accidental continuous
degeneracy appearing in previous mean field calculations and
result in two distinct states with higher symmetries:\\ a) the
uniaxial spin nematic with $Z_2$-symmetry;\\ b) the biaxial spin
nematic with $Dih_4$ symmetry.

We consider $F=2$ atoms in optical lattices using the following
Hamiltonian:
\begin{eqnarray}
{\cal H}&=& \sum_{k}
\frac{a_L}{2}\left(\hat{\rho}^2_k-\hat{\rho}_k\right) +\frac{b_L}{2}
\left(\hat{\cal F}^2_k-6\hat{\rho}_k\right) + 5
c_L {\cal D}^\dagger_k {\cal D}^{~}_k \nonumber \\
&-&t_L\sum_{<kl>} (\psi^\dagger_{k,\alpha\beta}
\psi^{~}_{l,\beta\alpha} + h.c.) - \sum_k \mu \hat{\rho}_k.
 \label{Hamiltonian}
\end{eqnarray}
Here $k$ is the lattice site index and $<kl>$ are the nearest
neighbor sites, $\mu$ is the chemical potential and
$t_L$ is the one-particle hopping amplitude.
$\psi^\dagger$ is a traceless symmetric tensor operator
that has been introduced in a previous work\cite{Zhou06};
$a_L, b_L,c_L$ are three on-site coupling constants which are
determined from three two-body scattering lengths $a_F$, $F=0,2,4$ and
on-site orbitals.
Components $\psi^\dagger_{\alpha\beta}$, $\alpha,\beta=x,y,z$
are linear superpositions of
five spin-2 creation operators, $\psi^\dagger_{m_F}$, $m_F=0,\pm1,\pm 2$.
The number
operator $\hat{\rho}$,
the dimer or singlet pair creation operator
${\cal D}^\dagger$,
the total spin operator $\hat{F}_\alpha$ are
defined as
$\hat{\rho}=1/2 tr \psi^\dagger \psi$, ${\cal
D}^\dagger={1}/{\sqrt{40}}tr \psi^\dagger\psi^\dagger$,
$\hat{F}_\alpha=-i\epsilon_{\alpha\beta\gamma}
\psi^\dagger_{\beta\eta}\psi_{\eta\gamma}$.

The amplitude of a condensate is a traceless symmetric tensor,
$\tilde{\chi}_{\alpha\beta}=<\hat{\psi}^{~}_{k,\alpha\beta}> $. In
this condensate, atoms occupy a one-particle spin state
$|\tilde{\chi}>$ that is defined as a linear superposition of five 
$F=2$ hyperfine states  $|2, m_F>$

\begin{eqnarray}
&& |\tilde{\chi}>=\sum_{\alpha\beta, m_F} \tilde{\chi}_{\alpha\beta}
{\cal C}_{\alpha\beta,m_F} |2,m_F>, \nonumber \\
&& {\cal C}_{\alpha\beta,m_F}=\sqrt{\frac{15}{4\pi}}\int {d\Omega}
\big( n_\alpha  n_\beta -\frac{1}{3}\delta_{\alpha\beta}\big ) Y^*_{2,m_F}(\theta,\phi). 
\label{wf}
\end{eqnarray}
Here $n_\alpha$, $\alpha=x,y,z$ are components of a unit vector
${\bf n}(\theta,\phi)$; $n_x=\sin\theta\cos\phi$,
$n_y=\sin\theta\sin\phi$ and $n_z=\cos\theta$. $Y_{2,m_F}$,
$m_F=0,\pm 1, \pm 2$ are five spherical harmonics with $l=2$. The
mean field energy per site as a function of $\tilde{\chi}$ can be
obtained as
\begin{eqnarray}
&& E_{MF} =
\frac{a_L}{8}\textrm{Tr}(\tilde{\chi}^*\tilde{\chi})\textrm{Tr}(\tilde{\chi}^*\tilde{\chi})
+\frac{c_L}{8}\textrm{Tr}(\tilde{\chi}^*\tilde{\chi}^*)\textrm{Tr}(\tilde{\chi}\tilde{\chi}) \nonumber \\
&& +\frac{b_L}{4} \textrm{Tr} [\tilde{\chi}^*, \tilde{\chi} ]^2 -
zt_L\textrm{Tr}(\tilde{\chi}^*\tilde{\chi})-\frac{\mu}{2}\textrm{Tr}(\tilde{\chi}^*\tilde{\chi}),
\label{energy}
\end{eqnarray}
where $z$ is the coordination number. Minimization of $E$ with respect
to the tensor $\tilde{\chi}$ yields nematic, cyclic and ferromagnetic 
phases studied previously.
Particularly when $c_L < 0, c_L < 4 b_L$ (as for rubidium atoms ($^{87}Rb$)), minimization of energy in
Eq.(\ref{energy}) requires that $\tilde{\chi}$ be a real symmetric
tensor (up to a phase). An arbitrary solution $\tilde{\chi}$ can be
obtained by applying an $SO(3)$ rotation and a $U(1)$ gauge
transformation to a real diagonal matrix $\chi$,
\begin{eqnarray}
\tilde{\chi}=\sqrt{4M} e^{i\phi} {\cal{R}} \chi {\cal{R}}^{-1};
\label{MFSolution}
\end{eqnarray}
${\cal R}$ is an $SO(3)$ rotation matrix, and $\chi$ is a normalized
real diagonal traceless matrix\cite{Zhou06},

\begin{eqnarray}
\chi=\left(
\begin{array}{ccc}
\chi_{xx} & 0 & 0 \\
0 & \chi_{yy} & 0 \\
0 & 0 & \chi_{zz}
\end{array}
\right),~~\textrm{Tr}(\chi\chi)=\frac{1}{2}. \label{mf}
\end{eqnarray}
Finally, the value of chemical potential $\mu$ is

\begin{eqnarray}
\mu_L =\mu + 2zt_L = \left(a_L+c_L\right)M, &
M=\frac{1}{2}\textrm{Tr}(\tilde{\chi}^*\tilde{\chi}). \nonumber
\label{ChiIsReal}
\end{eqnarray}
$M=<\hat{\rho}_k>$ is the average number of atoms per lattice site.
The shifted chemical potential $\mu_L = \mu+2zt_L$ measured from the
bottom of the band depends on scattering lengths through its
dependence on $a_L, c_L$.

In the mean field approximation, all of the quantum spin
nematics specified by different diagonal matrices Eq.(\ref{mf}) have
the same energy, i.e.~are exactly degenerate. For instance, when
$2\chi_{xx}=2{\chi}_{yy} =-\chi_{zz}=1/\sqrt{3}$, up to an overall
$SO(3)$ rotation, all atoms are condensed in the hyperfine state
$|2,0>$. This choice of $\chi$ represents a uniaxial spin
nematic. When $\chi_{xx}=-\chi_{yy}=1/2$ and $\chi_{zz}=0$, the
atoms are condensed in the state $\frac{1}{\sqrt{2}}\left(|2, 2>+
|2,-2>\right)$. In general, consider the
parametrization of the matrix elements: ($\xi\in [0,2\pi]$  )

\begin{eqnarray}
\chi_{xx}=\frac{\sin(\xi-\frac{\pi}{6})}{\sqrt{3}},
\chi_{yy}=\frac{\sin(\xi-\frac{5\pi}{6})}{\sqrt{3}},
\chi_{zz} = \frac{\sin(\xi-\frac{9\pi}{6})}{\sqrt{3}}.\nonumber \\
\label{xi-definition}
\end{eqnarray}

To study  ZPQFs, we first examine the energy spectra of
collective modes. For this, we expand $\psi^\dagger$ about the
mean field  $\tilde{\chi}$,
\begin{eqnarray}
\hat{\psi}^\dagger_{k,\alpha\beta} = \sqrt{4M}\chi_{\alpha\beta}
(\xi) + \sum_{\nu} L^\nu_{\alpha\beta}(\xi)
\hat{\theta}^\dagger_{k,\nu }\label{Decomposition}
\end{eqnarray}
where the superscript $\nu=x,y,z,t,p$ labels zero point motions of
$\chi$ along five orthogonal directions: three $SO(3)$ spin modes
($x$-,$y$-,$z$-mode) for rotations about the $x$, $y$ and $z$
axes, respectively, a spin mode ($t$-mode) for the motion along
the unit circle of $\xi$, and a phase mode ($p$-mode) describing
fluctuations of the condensate's overall phase. The corresponding
matrices are
\begin{eqnarray}
L^x=\left(
\begin{array}{ccc}
0 & 0 & 0 \\
0 & 0 & 1 \\
0 & 1 & 0
\end{array}\right),
L^y=\left(
\begin{array}{ccc}
0 & 0 & 1 \\
0 & 0 & 0 \\
1 & 0 & 0
\end{array}
\right), L^z=\left(
\begin{array}{ccc}
0 & 1 & 0 \\
1 & 0 & 0 \\
0 & 0 & 0
\end{array}\right)
\end{eqnarray}
and $L^{t}= 2\chi(\xi+\frac{\pi}{2})$, $~~L^{p}=2 \chi (\xi)$.
These matrices are multually orthogonal, $\textrm{Tr}(L^\mu
L^\nu)=2\delta_{\mu \nu}$. The five modes are decoupled and the
corresponding operators obey the bosonic commutation
relations, $\left[ \theta^{~}_{k,\mu},
\theta^{\dagger}_{l,\nu}\right] = \delta_{kl}\delta_{\mu\nu}$.

We expand the Hamiltonian in Eq.~(\ref{Hamiltonian}) using
Eq.~(\ref{Decomposition}) and keep the lowest order nonvanishing
terms. The result is a Hamiltonian for the fluctuations which is
bilinear in $\theta^\dagger_{k,\nu}$, $\theta_{k,\nu}$.  It can be
diagonalized by a Bogoliubov transformation. The result can be be
expressed in terms of Boguliubov operators,
$\tilde{\theta}^{\dagger}_{\mathbf{q},\nu}$ and
$\tilde{\theta}^{~}_{\mathbf{q},\nu}$,
\begin{eqnarray}
{\cal H} &=& \sum_{\mathbf{q},\nu} \sqrt{\epsilon_{\mathbf{q}} (2
m_{BN} v_\nu^2+  \epsilon_{\mathbf{q}})} \left(
\tilde{\theta}^{\dagger}_{\mathbf{q},\nu}
\tilde{\theta}^{~}_{\mathbf{q},\nu} + \frac{1}{2}\right).
\label{MFSpectrum}
\end{eqnarray}
Here $\epsilon_{\bf q}=4 t_L \sum_\alpha (1-\cos q_\alpha d_L )$
is the kinetic energy of an atom with crystal quasi-momentum ${\bf
q}=(q_x, q_y, q_z)$; $d_L$ is the lattice constant. $m_{BN}=1/4t_L
d_L^2$ is the effective band mass. $v_\nu$ ($\nu=x,y,z,t,p$) is
the sound velocity of the $\nu$-mode in the small-$|{\bf q}|$
limit;
$v^2_{\alpha=x,y,z}={4Mt_Ld_L^2\left(b_LG^{\alpha\alpha}-c_L\right)}$,
$v^2_t={4Mt_Ld_L^2\left(-c_L\right)}$,
$v^2_p=4Mt_Ld_L^2\left(a_L+c_L\right)$. The velocities of three
spin modes depend on a $\tilde{\chi}$-dependent $3\times 3$
symmetric matrix $G^{\alpha\beta}$,
$G^{\alpha\beta}=-(1/2)\textrm{Tr}\left(\left[L^\alpha,
L^p\right]\left[L^{\beta}, L^p\right]\right)$. More explicitly,
$G^{\alpha\beta}$ is a diagonal matrix with elements: $G^{xx}= 4
\sin^2 (\xi-\frac{2\pi}{3})$, $G^{yy}= 4 \sin^2
(\xi+\frac{2\pi}{3})$, $G^{zz}= 4 \sin^2 \xi $. Notice that the
velocity of the phase mode $v_p$ can be written as
$\sqrt{\mu_L/m_{BN}}$, only depending on the band mass $m_{BN}$
and the chemical potential $\mu_L$; it is independent
of $\xi$.  

\begin{figure}
\includegraphics[width=.95\columnwidth]{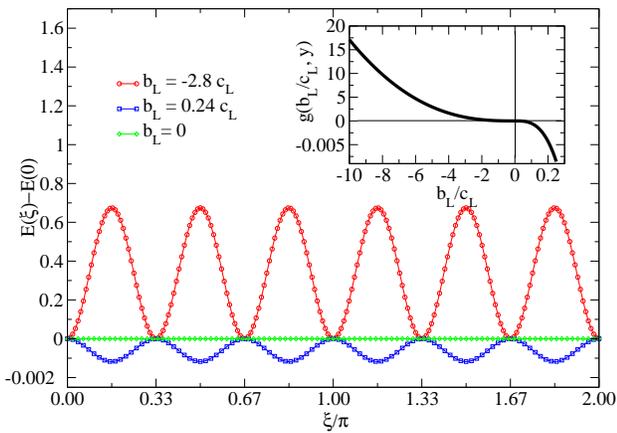}
\caption{(Color online) The zero point energy of spin nematics per lattice site as a function of $\xi$
(in units of $|M c_L|^{5/2}/t_L^{3/2}$, $c_L<0$).
Inset is $g(x,y)$, the amplitude of $E(\xi=\frac{\pi}{6})-E(0)$ as a 
function of $x(=b_L/c_L)$.
$y(=|c_L|/t_L)$ has been set to be $8\times 10^{-3}$.
A quantum first order phase transition occurs
at $b_L=0$ as a result of ZPQFs. For a positive $b_L$,
the energy minima correspond to uniaxial spin nematics while the maxima to biaxial spin nematics with
$Dih_4$ symmetries. }
\label{fig-energy}
\end{figure}

\begin{figure}
\includegraphics[width=\columnwidth]{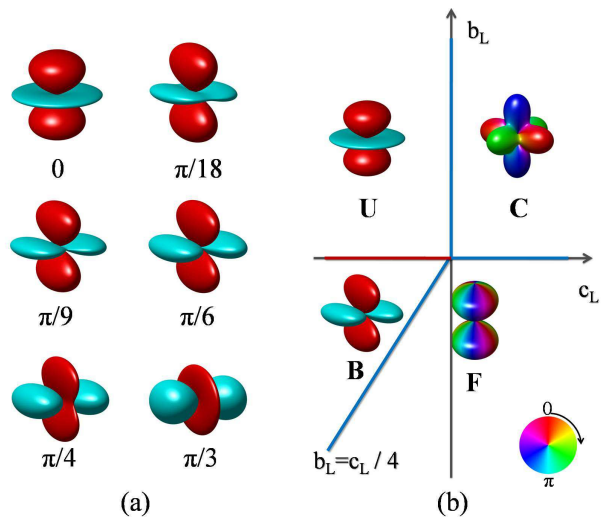}
\caption{ (Color online) (a)
Wavefunctions of spin nematics at various $\xi$; (b)
wavefunctions of uniaxial spin nematics (U), 
biaxial spin nematics with $Dih_4$ symmetries (B),
ferromagnetic condensates (F) and cyclic condensates (C). Plotted here are
spin wavefunctions $\Psi_S$ (represented by the usual
spherical harmonics) in spherical coordinates $(\rho, \theta,
\phi)$. $\rho=|\Psi_S(\theta,\phi)|$, and $\Psi_S
=\sum_{\alpha\beta} \sqrt{\frac{15}{4\pi}}\chi_{\alpha\beta}
n_\alpha n_\beta$. $n_\alpha$, $\alpha=x,y,z$ are components
of a unit vector ${\bf n}(\theta,\phi)$(see also
Eq.\ref{wf}); colors indicate the phase of wavefunctions.
The phase boundaries between U and C, C and F, F and B are obtained by 
minimizing Eq.(\ref{energy}) 
(similar to the calculations in Ref.\cite{Zhou06}); the phase 
boundary between U and B is obtained by 
taking into account ZPQFs.} 
\label{fig-wavefunction}
\end{figure}

Unlike the mean field energy $E_{MF}$ that is independent of
$\xi$, the zero point energy of spin nematics per lattice site $E$ is
$\xi$-dependent ( $\alpha$ summed over $x,y,z$),
\begin{eqnarray}
E(\xi) &=& \frac{1}{2 N_T}\sum_{\mathbf{q},\alpha}
\sqrt{\epsilon_{\mathbf{q}} (2m_{BN} v^2_{\alpha}(\xi)+
\epsilon_{\mathbf{q}})}. \label{FreeEnergy-simpleview}
\end{eqnarray}
Here $N_T$ is the number of lattice sites.
The main contribution to the $\xi$-dependence of energy is from 
fluctuations of wave vector
$|{\bf q}| \sim m_{BN} v_\alpha(\xi)$; for condensates, 
this characteristic momentum is
much smaller than $\hbar/d_L$.
The $\xi$-dependent energy is proportional to 
$\sum_\alpha v_\alpha^5/(d^5_L t^4_L) $;
the amplitude of it can be expressed as 
$|M c_L|^{5/2}/t_L^{3/2} g({b_L}/{c_L}, {|c_L|}/{t_L})$, where $g(x,y)$ 
is a dimensionless function that is studied numerically.
In Fig.~\ref{fig-energy}, we plot the energy as a function of
$\xi$. We find that there are two distinct phases when the
parameters occur in the region  $c_L <0$ and $b_L
>c_L/4$. They are separated by $b_L=0$ line. When $b_L$ is positive, $\xi =
0, \pi/3, 2\pi/3,...$  are ground states. These are uniaxial
nematic states which are rotationally invariant along an easy axis
(see Fig.~(\ref{fig-wavefunction} a)). When $b_L$
is negative, the points of $\xi= \pi/6, \pi/2, 5\pi/6,...$ are the
stable ground states while the uniaxial nematics are unstable.

The energy as a function of $\xi$ has following symmetries
\begin{eqnarray}
E(\xi) = E(-\xi), E(\frac{\pi}{3}+\xi) = E(\xi),
\end{eqnarray}
These are a result of the rotation and gauge invariance of the
energy function. They are found by examining the mean field
solutions in Eq.(\ref{MFSolution}),(\ref{xi-definition}). A
rotation ${\cal R}$ of $120^0$ around the $x=y=z$ line effectively
transforms a solution at $\xi$ to $\xi+ \pi/3$, or ${\cal R}^T
\chi(\xi) {\cal R}$= $-\chi(\xi+\frac{\pi}{3})$. And because the
energy is invariant under $SO(3)$ rotations and is an even
function of $\chi$, one finds that $E(\xi) = E(\xi+\pi/3)$. In
addition, a rotation of $180^0$ around the $x+y=0$ line in the
$xy$-plane transforms $\chi(\xi)$ to $\chi(-\xi)$, i.e. ${\cal
R}^T \chi(\xi) {\cal R}=\chi(-\xi)$; so similarly one finds that
$E(\xi)=E(-\xi)$. These symmetries only depend on rotational and
gauge invariance and are exact. So the energy is an even and
periodical function of $\xi$, with the period equal to $\pi/3$.
For an analytical function with these symmetries, $\xi=0$, $\pi/6$
and $\pi/3$ are always extrema.  Our calculations of the zero
point energy are consistent with this. The above observation also
indicates that up to an $SO(3)$ rotation and a phase factor,
states at $\xi$ and at $\xi+\pi/3$ are equivalent.

We now examine the topology of the manifold of ZPQF-induced
spin nematics. Without loss of generality, we first consider the
uniaxial nematic state at $\xi=0$, i.e.,
$\chi_{xx}=\chi_{yy}=-1/2\sqrt{3}$ and $\chi_{zz}=1/\sqrt{3}$; in
this case, the nematic easy axis specified by a unit vector ${\bf
e}$ is pointing along the $z$-direction (See Fig.
~\ref{fig-wavefunction} (a)). Such a uniaxial state is invariant
under an arbitrary rotation around the nematic axis ${\bf e}$; it
is further invariant under an inversion of the nematic axis ${\bf
e}\rightarrow -{\bf e}$ which is evident in the plot for the
wavefunction. The vacuum manifold for the uniaxial nematic is
therefore simply $S^2/Z_2 \otimes S^1$; here $S^2/Z_2$ is where
the nematic director lives and $S^1$ is the unit circle of
condensate phase variable. Unlike in uniaxial spin nematics of
$F=1$ atoms \cite{Zhou01}, here the spin orientation and
condensate phase are not entangled.

For the biaxial nematic state at $\xi=\pi/2$,
$\chi_{xx}=-\chi_{yy}=1/2$ and $\chi_{zz}=0$. The state is
invariant under the eight element {\em dihedral}-four group
$Dih_4$. The seven rotations that leave the state invariant (up to
a phase shift) are $90^0$, $180^0$ and $270^0$ rotations about the
$z$-axis, $180^0$ rotations about the $x$- and $y$-axes and about
the $x\pm y=0$ lines in the $xy$-plane. Four of these rotations,
the $90^0$, $270^0$ rotations around $z$-axis, and $180^0$
rotation around $x\pm y=0$ lines must be accompanied by a shift of
the phase of the condensate by $\pi$. The manifold therefore is
$[SO(3)\times S^{1}]/Dih_4$, where the Dihedral elements in the
denominator contain the rotations mentioned plus the $\pi$-phase
shifts. These nematics contain half-vortices.

On the other hand, a generic biaxial spin nematic (with no
accidental symmetries) is only invariant under a rotation of
$180^0$ around $x$-, or $y$- or $z$-axis and the invariant
subgroup is the Klein-four or dihedral-two group\cite{Semenoff06};
it has lower symmetries than either of the spin nematics selected
by ZPQFs.

The perturbative calculation carried out in this Letter is valid
when fluctuations of the order parameter $\chi$ are small. Here we
estimate the fluctuations along each of the five directions
$\nu=x,y,z,t,p$. Using Eq.~\ref{Decomposition}, we evaluate the  
amplitude of local fluctuations in an orthogonal mode, $A_1(\nu)=   
\langle 0|\theta^{\dagger}_{k,\nu}\theta^{~}_{k,\nu}|0 \rangle$. 
Taking into account the expression for $v_{\nu}$, 
we have the following estimate for the relative amplitude of
fluctuations, 

\begin{eqnarray}
\frac{A_1(\nu)}{M} \sim M^{{1}/{2}}
\big( \frac{\tilde{a}_\nu}{\tilde{a}_p} 
\frac{a_L+c_L}{t_L}\big)^{{3}/{2}}.
\end{eqnarray}
$\tilde{a}_\nu$ is the effective scattering length of each mode
which is a linear combination of scattering lengths in $F=0,2,4$
channels: $\tilde{a}_{\alpha} =
\frac{2-G^{\alpha\alpha}}{7}\left(a_2-a_4\right)-
\frac{1}{5}\left(a_0-a_4\right)$, $\tilde{a}_{t} =
\frac{2}{7}\left(a_2-a_4\right)- \frac{1}{5}\left(a_0-a_4\right)$,
$\tilde{a}_{p} =
\frac{1}{7}\left(2a_2+5a_4\right)+\frac{1}{5}\left(a_0-a_4\right)$.
For $^{87}Rb$ atoms in optical lattices, $\tilde{a}_{\alpha}/\tilde{a}_p 
\sim 10^{-2}$, $a_L \sim 50 nk$ and when $t_L \sim 300nk$, the relative 
amplitude of fluctuations
in $x$-, $y$-, $z$-, or $t$-mode is typically less than one percent. 
Finally, we would
like to point out that in nematic insulating states the properties
spin modes are very similar to what are
discussed here.

In conclusion, we have found that ZPQFs lift a continuous
degeneracy in spin nematics. Only uniaxial spin nematics or
biaxial nematics with dihedral-four symmetries are selected as the
true ground states. For $Rb^{87}$ atoms in the hyperfine spin-2
manifold, $b_L$ is between $-3c_L$ and $-10c_L$. According to the 
above analysis, the ground state should be a uniaxial spin nematic. The
ZPQF-induced energy landscape can be experimentally mapped out by
investigating the {\em macroscopic quantum dynamics} of
condensates prepared in certain initial states. A condensate of
rubidium atoms initially prepared at state
$\frac{1}{\sqrt{2}}(|2,2>+|2,-2>)$ (corresponding to $\xi=\pi/2$
point), because of the ZPQF-induced potential shown in
Fig.\ref{fig-energy}, could evolve towards a condensate with atoms
at state $\frac{1}{2}|2,0>
+\frac{\sqrt{3}}{2\sqrt{2}}(|2,2>+|2,-2>)$ (corresponding to
$\xi=\pi/3$ point). This leads to a temporal oscillation of
the population of atoms at state $|2,0>$, which can be studied in
experiments.  
This study could yield information on the potential.
We thank I. Bloch, K. Madison and E. Demler for discussions. This
work is in part supported by the office of the Dean of Science,
University of British Columbia, NSERC(Canada), Canadian Institute for Advanced 
Research, and the Alfred P.
Sloan foundation. 

{\em Note added}: After the submission of the manuscript, we notice that 
similar results are obtained in Ref.\cite{Turner07}.

\end{document}